\documentclass[12pt]{elsart}
\usepackage[english]{babel} 
\usepackage{epsfig}
\usepackage{changebar}
\usepackage{graphicx}
\begin{document}
\renewcommand{\figurename}{{\bf Fig.}}
\renewcommand{\tablename}{{\bf Tab.}}
\begin{frontmatter}
%
%
\title{A universal description for the freezeout parameters in heavy-ion
  collisions} 

\author{A.~Tawfik\thanksref{label2}}
\thanks[label2]{tawfik@physik.uni-bielefeld.de} 
\address{\small University of Bielefeld, P.O.~Box 100131, D-33501~Bielefeld, 
              Germany} 

\date{}

\begin{abstract}
It is shown that the freezeout parameters estimated in the heavy-ion collisions
all are well described by a constant value of the entropy density $s$ 
divided by $T^3$. The value of $s/T^3$ has been taken from the lattice QCD
simulations at zero baryo-chemical potential $\mu_B$ and assumed to remain
constant for finite $\mu_B$. This implies that the hadronic matter in rest
frame of produced system can be determined by constant degrees of
freedom. Furthermore, this condition has been used to predict
the freezeout parameters at low temperatures.   
\end{abstract}

\begin{keyword}
\PACS 12.40.Ee \sep 12.40.Yx \sep 05.70.Ce 
\end{keyword}


\end{frontmatter}

\section{\label{sec:1}Introduction}

At high temperature and/or density, it is conjectured that the confined
hadronic matter likely dissolves into quark-gluon plasma
(QGP)~\cite{Satz:1994uf}. Reducing the temperature, the QGP will
hadronize. At some temperature $T_{ch}$, the produced hadrons entirely
freeze out. $T_{ch}$ is the temperature at the chemical freezeout. The
freezeout parameters, $\mu_B$ and $T_{ch}$ can be determined in the thermal
models by combining various ratios of integrated particle
yields~\cite{Sollfrank:1999cy,Cleymans:1998yf}. Such a way, we obtain a
window in the $T_{ch}-\mu_B$ phase-diagram compatible to the 
experimental values. Both $\mu_B$ and $T$ are free parameters in these
thermal fits. The search for common properties of the freezeout
parameters in heavy-ion collisions has a long 
tradition~\cite{Cleymans:1999st,Braun-Munzinger:1996mq}. Different models
have been suggested, for example the average energy per average particle number
$<E>/<n>=1~$GeV~\cite{Cleymans:1999st}, total baryon number density
$n_b+n_{\bar{b}}=0.12~$fm$^{-3}$~\cite{Braun-Munzinger:1996mq} and the
interconnection amongst the constituents of the hadronic
matter~\cite{Magas:2003wi}. In this article, we assume that one best
achieves this objective by assigning the 
entropy density $s$ normalized to $T^3$ to a constant
value~\cite{Tawfik:2004ss}. The results can be summarized in the following
way: the hadronic matter in rest frame of produced system can be 
determined by constant degrees of freedom\footnote[1]{$\pi^2/4$-scaling of
  $s/T^3$ gives the effective degrees of freedom of free 
  gas~\cite{Cleymans:2005km}}. 
For vanishing free energy, i.e. at the chemical freezeout, the {\it
  equilibrium} entropy gives the amount of energy which can't be used to
  produce additional work. We can in this context define the 
entropy as the degree of sharing and spreading the energy inside the
equilibrium system. Furthermore, we find    
that the strangeness degrees of freedom are essential at low collision
energies, where the strangeness chemical potential $\mu_S$ is as large as
$\mu_B$. According to the strangeness conservation in the heavy-ion
  collisions, we find that the higher is the collision energy, the smaller is
$\mu_S$~\cite{Tawfik:2004sw}.

\section{\label{sec:2}The model}

In order to describe the freezeout phase-diagram, we basically have to be
able to determine two characteristic regions. The first one is at
\hbox{$\mu_B=0$} and finite $T_{ch}$. The second one is characterized by
$T_{ch}=0$ and finite $\mu_B$. Localizing the first region has been the 
subject of different experimental studies~\cite{Becattini:2002nv}. It has
been found that \hbox{$T_{ch}(\mu_B=0)\approx
  174\;$MeV}~\cite{Braun-Munzinger:2003zz}. The 
lattice estimation for the deconfinement temperature gives
\hbox{$T_{c}(\mu_B=0)=173\pm8\;$MeV}~\cite{Karsch:2001cy}. This  
implies that the deconfinement and freezeout seem to be coincident at
small $\mu_B$. For the second region, we are left with effective
models. As $T$ is close to zero, one expects that the nucleons will
dominate the resonance gas. We find that at $T=0$ the
baryo-chemical potential is corresponding to the normal nuclear density, 
\hbox{$n_0\approx0.17\;$fm$^{-3}$}, i.e. 
\hbox{$\mu_{B}\approx0.979\;$GeV}. We have to mention here that this value
can slightly be different according to the initial
conditions~\cite{Fraga:2003uh}. As we shall see later, extrapolating our
freezeout curve to the abscissa might result in a $\mu_B$-value very close
to this estimation. But it is important to notice that the
condition of constant $s/T^3$ breaks up at $T=0$.

The partition function in the hadronic matter at finite $T$, $\mu_B$ and
$\mu_S$ is given by the contributions of all hadron resonances up to
$2~$GeV treated as a free quantum  
gas~\cite{Karsch:2003vd,Karsch:2003zq,Redlich:2004gp,Tawfik:2004sw,Tawfik:2005qh}   
\begin{eqnarray}
\ln {Z}(T,\mu_B,\mu_S) &=& V\,\frac{g}{2\pi^2} \int_{0}^{\infty}
           k^2 dk  \ln\left[1 \pm\, \gamma\,
           \exp{\frac{\mu_B+\mu_S-\varepsilon}{T}}\right],
           \label{eq:lnz} 
\end{eqnarray}
where $\varepsilon=(k^2+m^2)^{1/2}$ is the single-particle energy and 
$\pm$ stand for bosons and fermions, respectively. $g$ is the spin-isospin
degeneracy factor. $\mu_S$, the strangeness chemical potential, has been
calculated as a function of $T$ and $\mu_B$ under the condition that the
average strange particle number $<n_s>$ equals to the average anti-strange
particle number $<n_{\bar{s}}>$, the so-called strangeness
conservation. $\gamma\equiv\gamma_q^n\gamma_s^m$ is the quark phase space
occupancy parameter, with $n$ and $m$ being the number of light and strange
quarks, respectively. In carrying out our calculations, we used
$\gamma_q=\gamma_s=1$, i.e. we assumed that the phase-space occupancy
parameters of both light and strange quarks are in
equilibrium~\cite{Rafelski:1991rh,Tawfik:2005gk} and the particle
production is due to a chemically equilibrium process. 

At $T=0$, the hadron resonance gas model (HRGM) is no longer
applicable. The laws of thermodynamics can't hold in this
limit. Corrections due to van~der~Waals repulsive interactions have not
been taken into account in our calculations~\cite{Tawfik:2004sw}. 

\section{Results}

\begin{figure}[thb] 
\centerline{\includegraphics[width=12.cm]{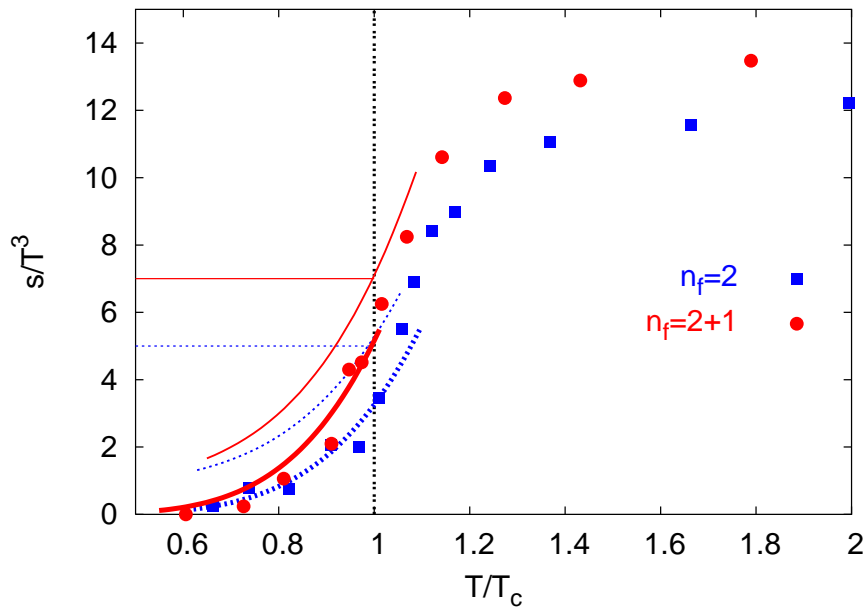}} 
\caption{\label{Fig:3}\footnotesize Lattice QCD results on the entropy
  density normalized to $T^3$ for $n_f=2$ (full circles) and $n_f=2+1$
  (full squares) quark flavors at 
$\mu_B=0$~\cite{Karsch:2000kv,Karsch:2003zq} on the top of the
results from the hadron resonance gas model (curves). The thick curves
represent the results from hadron resonance gas model (HRGM) with rescaled
masses. We find a well agreement with the lattice QCD simulations. The HRGM 
calculations with the physical resonance masses are given by the thin
curves. The horizontal lines indicate to $s/T^3$-values for the physical
resonance masses at the different quark flavors and critical temperature
$T_c$. 
}    
\end{figure}

\begin{figure}[thb] 
\centerline{\includegraphics[width=12.cm]{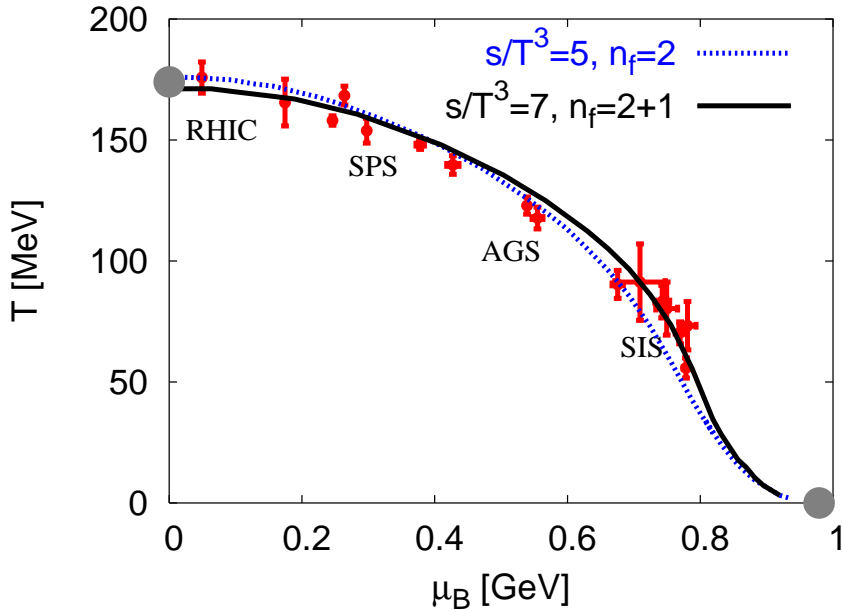}} 
\caption{\label{Fig:4}\footnotesize
  The freezeout parameters according to constant $s/T^3$ on the top of the
  experimentally estimated (or thermally fitted) freezeout parameters
  (small full circles with errors). For non-strange 
  hadron resonances, we use $s/T^3=5$ and for all hadron resonances we
  assign $s/T^3$ to $7$. Both values have been taken from lattice QCD
  simulations at zero baryo-chemical potential (Fig.~\ref{Fig:3}). We find
  that strangeness degrees of freedom are essential for reproducing the
  freezeout parameters at low incident energies. The condition of constant
  $s/T^3$ have been applied for $T>5\,$MeV. At smaller temperatures, the
  HRGM is no longer applicable.}
\end{figure}

We plot in Fig.~\ref{Fig:3} the lattice results on $s/T^3$ vs. $T/T_c$ at
$\mu_B=0$ for different quark flavors
$n_f$~\cite{Karsch:2000kv,Karsch:2003zq}. The quark masses used in the lattice
calculations are heavier than their physical masses in vacuum. For a
reliable comparison with the lattice QCD, the hadron resonance
masses included in HRGM have to be re-scaled to values heavier than the
physical ones~\cite{Karsch:2003vd,Karsch:2003zq}. As shown in
Fig.~\ref{Fig:3}, HRGM 
 can very well reproduce the lattice results for the different quark
 flavors under this re-scaling condition. In the same figure, we draw the
 results for physical resonance masses as thin curves, i.e. the case if
 lattice QCD simulations were done for physical quark masses. The two
 horizontal lines point at the values of $s/T^3$ at the critical temperature
 $T_c$. As mentioned above, the
 critical $T_c$ and the freezeout temperature $T_{ch}$ are
 assumed to be the same at small $\mu_B$. We find that
 $s/T^3=5$ for $n_f=2$ and $s/T^3=7$ for $n_f=2+1$. These
 two values will be used in order to describe the freezeout parameters for
 different quark flavors. The normalization with 
 respect to $T^3$ should not be connected 
with massless particles. Either the resonances in HRGM or the quarks on
lattice are massive. We divide $s$ by $T^3$ in order to remove the
$T$-dependence.

In HRGM, we calculate the temperature $T_{ch}$ at different $\mu_B$
according to constant $s/T^3$. $\mu_B$ is corresponding to the collision
energy. On the other hand, $\mu_S$, the strangeness chemical potential, has
been calculated in dependence on $\mu_B$ and $T$. The resulting $T$ and $\mu_B$
are plotted in Fig.~\ref{Fig:4} on the top of the freezeout parameters
($T_{ch}$ and $\mu_B$) which, as mentioned above, have been estimated by
thermal fits of various ratios of the particle yields produced in different
heavy-ion collisions. The dotted curve represents our results for
$n_f=2$. In this case, only the non-strange hadron resonances are included in
the partition function, Eq.~(\ref{eq:lnz}). The entropy is given by
$\partial T\ln{Z}(T,\mu_B,\mu_S)/\partial T$. The condition applied  
in this case is $s/T^3=5$. The solid curve gives $2+1~$results (two light
quarks plus one heavy strange quark). Here all resonances are included in
the partition function and the condition reads $s/T^3=7$. We find that both
conditions can satisfactory describe the freezeout parameters at high
collision energy. We notice that the $n_f=2$ curve doesn't go through
the SIS data points. Therefore, we can conclude that the non-strange
degrees of freedom alone might be non-sufficient at the SIS energy.

We have to mention here that both the entropy density $s$ and the corresponding
temperature $T_{ch}$ decrease with increasing $\mu_B$. The entropy density
is much faster than $T$, so that the ratio $s/T^3$ becomes greater than $7$
at very large $\mu_B$. In this limit, {\it thermal} entropy density $s$ is
expected to vanish\footnote[2]{At $T=0$ the thermal entropy equals to zero
  as the system gets degenerate.}, since it becomes proportional to $T$
(third law of thermodynamics). The {\it quantum}
entropy~\cite{Miller:2003ha,Miller:2003hh,Miller:2003ch,Miller:2004uc,Hamieh:2004ni,Miller:2004em} is entirely disregarded in these calculations.

\section{\label{sec:4}Conclusion and outlook}

We used HRGM in order to map out the freezeout curve according to constant
$s/T^3$. Taking its value from the lattice QCD simulations at $\mu_B=0$ and
assuming that it remains constant in the entire $\mu_B$-axis, we obtained
the results shown in Fig.~\ref{Fig:4}. We find that the freezeout
parameters $T_{ch}$ and $\mu_B$ are very well described under this
condition. We also find that the strangeness
degrees of freedom seem to be essential at low collision energies, where the
strangeness chemical potential $\mu_S$ is as large as
$\mu_B$. At high collision energies, $\mu_B$ decreases and correspondingly
$\mu_S$. We conclude that the given
ratio $s/T^3$ characterizes very well the final states observed in all
heavy-ion collisions. The hadronic abundances observed in the final state of
heavy-ion collisions are settled when $s/T^3$ drops to $7$, i.e. 
the degrees of freedom drop to $7\,\pi^2/4$. Meanwhile the changing in the
particle number with the changing in the collision energy is given by the
baryo-chemical potential 
$\mu_B$, the energy that produces no additional work, i.e. the stage of
vanishing free energy, gives the entropy at the chemical 
equilibrium. At the chemical freezeout, the equilibrium entropy  
represents the amount of energy that can't be used to produce additional
work. In this context, the entropy is defined as the degree of sharing and
spreading the energy inside the system that is in chemical equilibrium.  

Regarding to the sharp peak of the $K^+/\pi^+$ ratio at SPS
energy~\cite{Gazdzicki:1998vd}, we still face 
the problem that the thermal models (HRGM), on the one hand side, describe
very well the freezeout parameters at each collision energy individually,
i.e. they are assumed to be able to reproduce all particle ratios. On the
other hand, we find that such a sharp peak can't be reproduced by the 
thermal models~\cite{Cleymans:2004hj}. Indeed, we find that the $K^+/\pi^+$
ratio at the SIS, AGS and low SPS energies can be reproduced by the thermal
models. But a large overestimation is to be observe at top SPS and all
RHIC energies. If we assume that the uncertainty in the experimentally
estimated (or fitted by the thermal models) $K^+/\pi^+$ ratio is small,
then one might think that this experimental observation might indicate to
some critical phenomena, which we don't include in the thermal models. 

One way to avoid this dilemma might be assuming two separate conditions for the
freezeout parameters. One of them is to be applied at small $\mu_B$, where
we can't distinguish between the freezeout and the deconfinement. The
other 
one should be able to describe the freezeout parameters at large $\mu_B$. Both
conditions should be able to reproduce the particle ratios at all collision
energies. The region, i.e. $\mu_B$-values, where the two conditions 
intersect hopefully will be corresponding to the collision
energy, at which the sharp peak of the $K^+/\pi^+$ ratio has been
observed. 

In~\cite{Tawfik:2005gk}, we have worked out an additional model. We have
allowed the phase space occupancy parameters $\gamma_q$ and $\gamma_s$ to take
values other than that of equilibrium.


\end{document}